\documentclass[twocolumn,tighten,twocolappendix]{aastex702}

\usepackage{apjfonts}
\usepackage{amssymb}
\usepackage{amsmath}

\usepackage{graphicx}
\usepackage{multirow}
\usepackage{enumitem}

\usepackage{color}
\usepackage[dvipsnames]{xcolor} 
\usepackage{soul}
\usepackage[normalem]{ulem}

\usepackage{xspace}
\usepackage{xifthen}
\usepackage{changepage}

\usepackage{listings}
\definecolor{dkgreen}{rgb}{0,0.6,0}
\definecolor{gray}{rgb}{0.5,0.5,0.5}
\definecolor{mauve}{rgb}{0.58,0,0.82}
\definecolor{golden}{rgb}{0.86,0.65,0.01}
\newcommand{\code}[1]{\texttt{#1}\xspace}

\begin{document}
\shorttitle{Products of AGB Mass Transfer in Stellar Streams}
\shortauthors{Weatherford, McKenzie, \& Usman}

\title{AGB-star Mass Transfer Products in Stellar Streams as Tracers of Density in Dissolved Star Clusters}
\linenumbers

\author[0000-0002-9660-9085]{Newlin C. Weatherford}
\affil{Observatories of the Carnegie Institution of Washington, 813 Santa Barbara Street, Pasadena, 91101, CA, USA}
\email{nweatherford@carnegiescience.edu}

\author[0000-0002-1715-1257]{Madeleine McKenzie}\thanks{NASA Hubble Fellow}
\affil{Observatories of the Carnegie Institution of Washington, 813 Santa Barbara Street, Pasadena, 91101, CA, USA}
\email{mmckenzie@carnegiescience.edu}

\author[0000-0003-0918-7185]{Sam A. Usman}
\affil{Colgate University, 13 Oak Drive, Hamilton, NY 13346, USA\vspace{-0.1cm}}
\email{susman@colgate.edu}

\begin{abstract}
The least dense globular clusters (GCs) appear to host the highest fractions of stars with chemical signatures of mass transfer from asymptotic giant branch (AGB) companions. Recent observations suggest the fraction may be even higher in stellar streams that are debris of tidally disrupted (preferentially low-density) GCs. We explore whether this is attributable to lower encounter rates in less-dense GCs allowing more of the relevant binaries to survive dynamical hardening, or outright disruption, long enough to undergo AGB mass transfer. We run a small suite of GC simulations varying initial density using the code \texttt{CMC} with its latest observationally motivated binary prescriptions. Using \texttt{CMC}'s built-in \texttt{COSMIC} code for binary stellar evolution, we first identify primordial binaries that would, if left unperturbed, likely undergo AGB mass transfer via either standard or wind Roche-lobe overflow. We then check how many of those binaries survive GC dynamics in \texttt{CMC} long enough to experience AGB mass transfer. We find that wind Roche-lobe overflow is necessary to reproduce the fraction of (sub)giant stars in Galactic GCs that are likely products of AGB mass transfer (${\sim}1\%$) and that binary dynamics plausibly explain why this fraction appears lower in dense, intact GCs than stellar streams. The abundance of mass-transfer products (also collision products and binaries) in upcoming surveys of stellar streams without known progenitors will thus probe the densities of dissolved GCs---difficult to determine otherwise but key to improved stream modeling and predicted rates of many astrophysical transients.
\vspace{-0.15cm}
\end{abstract}


\section{Introduction} \label{sec:intro}
As some of the densest stellar environments, globular clusters (GCs) are essential factories of stellar exotica and energetic transients arising from close stellar encounters, including X-ray binaries \citep{Clark1975, Angelini2001, Pooley2003, Kremer2018_XRBs}, radio pulsars \citep{Lyne1987, Sigurdsson1995, Lorimer2008, Ye2019}, fast radio bursts \citep{Kremer2021, Kirsten2022}, gravitational-wave sources \citep{Rodriguez2015, Rodriguez2021, Askar2017}, tidal disruption events \citep{Kremer2019_TDE}, and stellar collisions that may grow intermediate-mass black holes \citep{PortegiesZwart2002, Giersz2015, GonzalezPrieto2024, Haberle2024}. How much GCs contribute to these phenomena depends on their birth densities, which GCs today may misrepresent, being only the survivors of an initially larger population dissolved by internal dynamics and the Galactic tide \citep{Vesperini1998, FallZhang2001}. Diffuse systems are more vulnerable to tidal stripping, so surviving GCs may be those born the most dense, but the opposite may be true if the densest GCs formed under the strongest tides \citep{Gieles2023}. Clues may exist in our Galaxy's many stellar streams found to be debris of dissolved GCs \citep{BonacaPriceWhelan2025}. Progenitor density is already known to affect stream kinematics relevant to understanding dark matter \citep{Weatherford2026, Phillips2026}. The contents of streams may in turn provide insight into the progenitors.

Spectroscopic surveys of GCs and streams suggest a density dependence in the abundance of stars with chemical signatures of mass transfer from asymptotic giant branch (AGB) companions. These AGB mass-transfer (AGB-MT) stars fall into several categories (e.g., Ba, CH, S-type, and CEMP\nobreakdash-s stars) enhanced in carbon and/or slow-neutron-capture ($s$\nobreakdash-process) elements they are not evolved enough to have produced themselves \citep{Iben1983, Jorissen2019}. Radial velocity measurements reveal most are binaries, as expected if their unseen companions are white-dwarf remnants of former AGB donors \citep{Lucatello2005, Hansen2016, Jorissen2019}. AGB-MT stars make up from ${\sim}1\%$ \citep{MacConnell1972, Luck1991, Buder2025} to $13\%$ \citep{Placco+2014, Arentsen+2022} of subgiant and red giant branch (RGB) stars in our Galaxy's halo but only ${\approx}0.4\%$ of those in GCs \cite{DOrazi2010}. Since their earliest detection in GCs, AGB-MT stars have also appeared rarest in the densest clusters \citep{McClure1977, Cote1997, DaCosta2025}, while recently appearing comparatively often in streams, including one of 8 RGB stars in 300S \citep{Usman2024} and one of 12 in GD-1 (\citealt{Zhao2025}; S.~A. Usman et al. in prep.). Accounting also for non-detections in Atlas/Aliqa-Uma and Phoenix \citep[out of 20 RGB stars examined;][]{Ji2020b}, this implies that AGB-MT stars may be ${\sim}10$ times more abundant in streams than intact GCs (S.~A. Usman et al. in prep.).

An AGB-MT star forms when a binary's more massive star reaches the thermally pulsing AGB phase, where repeated third dredge-up brings carbon and $s$-process elements to its surface, before transferring some of this material to its companion \citep{Busso+1999, Herwig+2005, Karakas_Lattanzio2014}. The rates of the nuclear reactions providing most slow neutrons depend on several parameters, including donor mass, which can thus be inferred by fitting relative abundances of first-peak (Sr, Y, Zr), second-peak (Ba, La, Ce), and third-peak (Pb) $s$-process elements to expected AGB yields \citep{Gallino+1998, Cristallo+2015, Karakas_Lugaro2016}. Such fits favor low-mass donors, of ${\approx}1.3$--$2\,M_\odot$ for CEMP-s stars \citep{Bisterzo+2011, Bisterzo+2012} and ${\approx}2$--$3\,M_\odot$ for Ba stars \citep{Cseh+2022, denHartogh+2023}, with a sample of 180 Ba stars implying donors below $4\,M_\odot$ \citep{Vilagos+2024}. The AGB-MT star in 300S likewise points to a $2.6\,M_\odot$ donor (G.~C. Cingquegrana et al. in prep.), while studies at lower metallicities more typical of GCs and stellar streams, ${\rm [Fe/H]}\sim -2$, suggest donors as small as $1\,M_\odot$ can produce the LMC's carbon stars \citep{Stancliffe2005}.

The donor can transfer mass in three ways. In standard Roche-lobe overflow (RLOF), it fills its Roche lobe, but since AGB stars have deep convective envelopes that rapidly expand with mass loss, RLOF onto a less massive companion is often unstable, leading instead to common-envelope evolution. In very wide orbits, the companion instead captures only a small fraction of the AGB wind via Bondi-Hoyle-Lyttleton (BHL) accretion. In between these regimes is wind Roche-lobe overflow \citep[WRLOF;][]{Mohamed_Podsiadlowski2007}. AGB winds are slow, driven by radiation pressure on dust that condenses a few stellar radii from the star. If this dust-formation radius expands beyond the donor's Roche lobe, the wind is funneled through the inner Lagrange point toward the companion, transferring mass far more efficiently than BHL accretion \citep{Chen+2017, Saladino+2019}. By enabling efficient mass transfer in orbits too wide for RLOF, WRLOF increases the expected frequency of AGB-MT stars \citep{Abate2013}.

The scarcity of AGB-MT stars in dense GCs likely reflects strong perturbations to these binaries before mass transfer can occur \citep{Cote1997, DaCosta2025}. Strong encounters in a GC's core can disrupt a binary, exchange one of its members, or soften or harden its orbit \citep[e.g.,][]{Heggie1975} enough that mass transfer is no longer possible or common-envelope evolution occurs instead. The widest binaries expected to survive in GCs for a Hubble time are similar in size to the widest isolated CH-star binaries, suggesting that hardening is chiefly responsible for the rarity of AGB-MT stars in GCs \citep{Cote1997}. The wide orbits that WRLOF makes efficient at mass transfer are especially vulnerable to hardening or disruption.

Similar logic may cause streams to host more AGB-MT stars than intact clusters. For GCs of typical or higher density, the evaporation rate (and so stream density) grows shallowly with GC density due to faster relaxation \citep{McLaughlin2008, Gieles2008, Weatherford2024}. Yet diffuse GCs are more vulnerable to tidal stripping, exacerbated by prolonged retention of black holes driving GC expansion, so their evaporation quickens with falling density \citep{Giersz2019, Gieles2023}. The least dense GCs should thus yield the most-prominent streams; indeed, Galactic GCs with streams or elevated mass-loss rates tend to be less dense than average \citep{Chen2026, Wang2026}. The correspondingly gentler dynamics should make AGB-MT stars more abundant in streams than most intact GCs.

In this Letter, we explore the origins of the observed density dependence of AGB-MT star abundance in GCs and streams and highlight the potential ramifications for mass-transfer physics and possible usage to infer initial densities of dissolved GCs. We describe our cluster simulations and methodology in Section~\ref{sec:simulations}, including calculations with and without rough prescriptions for WRLOF, and present our results in Section~\ref{sec:results}. We discuss caveats, implications for inferring progenitor density, and potential future applications in Section~\ref{sec:discussion}, and summarize our findings in Section~\ref{sec:conclusion}.

\section{Cluster Models} \label{sec:simulations}
We study the survival rate of binaries capable of AGB-MT versus cluster density in a small set of GC simulations using \code{Cluster Monte Carlo} \citep[\code{CMC};][]{CMCRelease}, a state-of-the-art implementation of the orbit-averaged Monte Carlo method \citep{Henon1971a, Henon1971b} in continuous use and development for the last 25 years. 
\code{CMC} capably handles the essential physics of binaries in GCs, including direct integration of strong binary--single and binary--binary encounters, and binary stellar evolution with \code{COSMIC} \citep{COSMIC}. 
We run five distinct GC simulations up to an age of $14\,{\rm Gyr}$ or dissolution, varying only the initial virial radius, $r_v/{\rm pc}=(1,2,4,8,16)$, while choosing typical values for other key parameters. We choose initial number of stars $N_0=8\times 10^5$, metallicity $Z=0.1\,Z_\odot$ ([Fe/H]=$-1$), and Galactocentric distance $R_{\rm gc}=8\,{\rm kpc}$. After ${\sim}10\,{\rm Gyr}$ of evolution, these choices yield models that closely match typical Milky Way GCs \citep{CMCCatalog}. The only GC that dissolves in ${<}14\,{\rm Gyr}$ is the one initialized with $r_v=16\,{\rm pc}$, dissolving after ${\approx}7\,{\rm Gyr}$.

We use \code{CMC}'s latest observationally motivated recipes for initial binaries \citep{OConnor2026b}, starting by drawing all star masses $m$ from the \cite{Kroupa2001} initial mass function spanning $m/M_\odot\in[0.08,150]$.
We pair stars into binaries in a way that closely reproduces the mass-dependent fraction $f_b$ of stars in the solar neighborhood with companions closer than $10\,\rm{AU}$ \citep[Table 1 of][]{Offner2023}. The pairing assumes uniformly distributed binary mass ratios $q\in[0.1,1]$, or $q\in[0.6,1]$ when the primary mass $m_1>15\,M_\odot$ \citep{DuquennoyMayor1991, Sana2012}. To avoid extrapolating beyond observational support, we truncate the resulting formula---$f_b(m_1) = 0.15 + 0.34\,\log_{10}(m_1/0.55\,M_\odot)$---to $f_b=0.15$ for $m_1<0.55\,M_\odot$ and $f_b=0.82$ for $m_1>50\,M_\odot$. Over ${\sim}10\,{\rm Gyr}$, the total $f_b$ in our models drops from ${\approx}17\%$ to ${\lesssim}10\%$, as in Galactic GCs today \citep{Milone2012, JiBregman2015}.

For $m_1\leq15\,M_\odot$, we sample binary eccentricity $e$ uniformly \citep[following \emph{Gaia} astrometry of close binaries;][]{Hwang2022} and draw semi-major axis $a$ from a log-normal distribution \citep{Raghavan2010} ranging from barely avoiding RLOF in the pre-main-sequence phase to $a_{\rm max}=10\,{\rm AU}$. So all simulations start with an equal number of binaries, $N_{\rm mt,iso}$, that would undergo AGB-MT if left unperturbed. (In fact, they all start from the \emph{same} set of stars and binaries, differing only in radial distribution.)
For $m_1\geq15\,M_\odot$, we instead sample $(a,e)$ from the distributions for O-type stars reported by \cite{Sana2012}.

Except for minor updates to \code{COSMIC} affecting compact objects, we adopt identical input stellar evolution physics as the \code{CMC} Cluster Catalog \citep{CMCCatalog}. \code{COSMIC} models standard RLOF but not WRLOF, which may form most AGB-MT stars \citep{Abate2013,Abate2015,Krynski2025}. So while we consider two different scenarios---one where all AGB-MT stars form via RLOF, and one where WRLOF contributes---our consideration of the latter possibility relies on a simple post-processing criterion. \code{COSMIC} automatically records when standard RLOF occurs, assumed to be once the donor's radius $R_1$ exceeds its Roche radius $R_{L,1}$ \citep[using the formula by][]{Eggleton1983}, but we manually check for onset of WRLOF every $0.1\,{\rm Myr}$ of each binary's evolution. We assume WRLOF would occur once $R_{\rm dust}>R_{L,1}$ \citep[e.g.,][]{Krynski2025}, estimating the donor's dust formation radius as \citep{Hofner2007}
\begin{equation} \label{eq:rdust}
R_{\rm dust} = \frac{1}{2}R_1\left(\frac{T_{\rm eff,1}}{T_{\rm dust}} \right)^{2.5},
\end{equation}
where $T_{\rm eff,1}$ is the donor's effective temperature and $T_{\rm dust}$ is the dust condensation temperature. $T_{\rm dust}\approx(1500\,{\rm K},1000\,{\rm K})$ for oxygen-rich and carbon-rich stars (C/O < 1, C/O > 1); we use these values for donors in the early AGB phase and thermally pulsing AGB (TP-AGB) phase, respectively.

In both scenarios, the fraction of initial binaries capable of AGB-MT that survive GC dynamics long enough for AGB-MT to occur is $f_{\rm surv}=N_{\rm mt}/N_{\rm mt,iso}$. To determine $N_{\rm mt,iso}$, we use \code{COSMIC} to evolve binaries for $t_{\rm evol}=11\,{\rm Gyr}$ as if isolated (unperturbed), and record how many start AGB-MT and when ($t_{\rm mt}$). We count only binaries satisfying the following:
\begin{itemize}
\item The donor's initial mass is $1$--$5\,M_\odot$ (see Section~\ref{sec:intro}).
\item Mass transfer starts during the donor's AGB or TP-AGB phases and ends in its TP-AGB phase or exit thereof (by becoming a white dwarf).
\item From $t_{\rm mt}$ to $t_{\rm evol}$, the binary does not undergo any other mass transfer or common envelope (assumed to prevent detectable enrichment of the accretor's surface).
\item At $t_{\rm evol}$, the accretor is a subgiant (Hertzsprung gap) or RGB star, like most observed AGB-MT stars.
\end{itemize}

We then determine $N_{\rm mt}$ by running the \code{CMC} simulations, starting again at $t=0$ with identical \code{COSMIC} settings, and identifying the first time $t_{\rm enc}$ each binary initially capable of AGB-MT is strongly perturbed by another star. The perturbation necessary to prevent AGB-MT is uncertain and depends on the specific trajectories and hydrodynamics involved. We conservatively assume any encounter prior to $t_{\rm mt}$ prevents detectable enrichment of the accretor if it changes $a$ or $e$ by ${>}10\%$; changes larger than this usually arise from resonance interactions \citep[instead of flybys;][]{HeggieHut1993}, which may disrupt mass transfer. We tested allowing only changes ${<}1\%$, but this decreased $f_{\rm surv}$ by only a few percent, so the exact cutoff does not affect our conclusions.

\begin{figure*}[pht!]
\centering
\includegraphics[width=\textwidth]{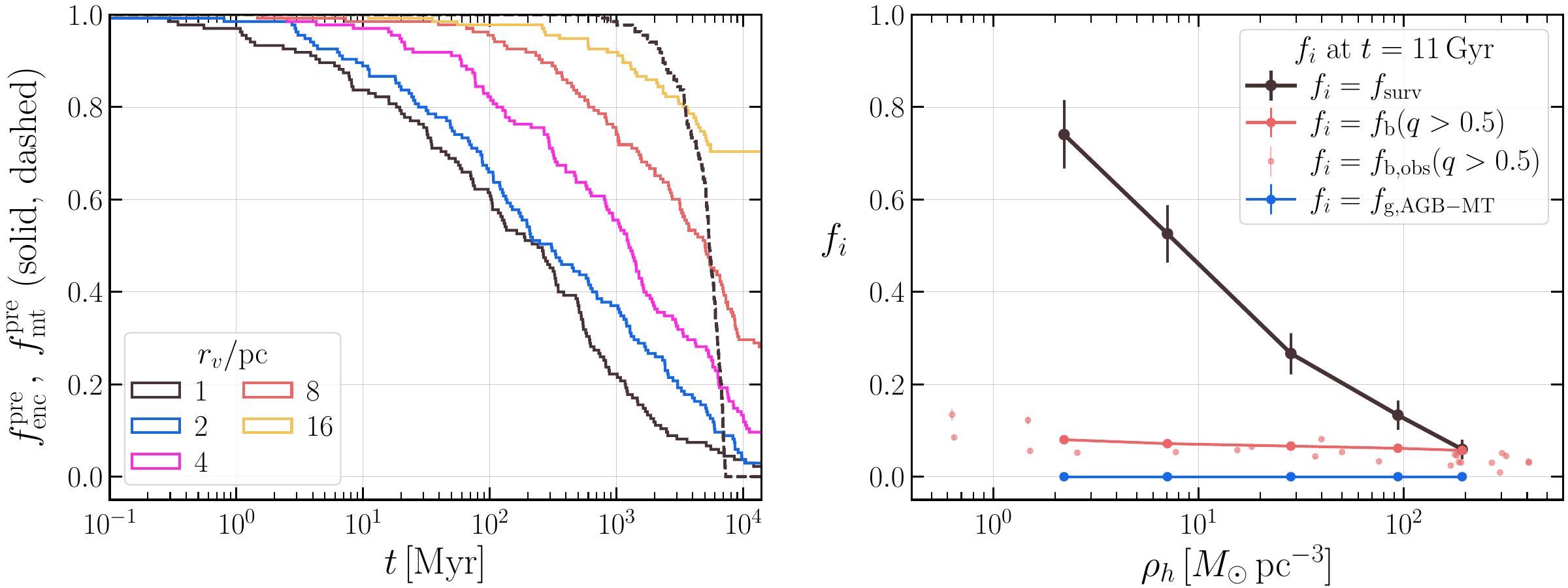}
\caption{Scenario 1: AGB-MT occurs only via standard RLOF. \emph{Left}: Time evolution in the fraction of a GC's binaries initially capable of AGB-MT that have yet to undergo a strongly perturbing encounter ($f^{\rm pre}_{\rm enc}$; solid curves) or initiate mass-transfer even if unperturbed ($f^{\rm pre}_{\rm mt}$; dashed curve). Color indicates the initial virial radius $r_v$ in each \code{CMC} simulation ($r_v\approx1.3\,r_{h,0}$, where $r_{h,0}$ is the initial half-mass radius). The crossing point of $f^{\rm pre}_{\rm mt}(t)$ and $f^{\rm pre}_{\rm enc}(t)$ closely approximates the fraction $f_{\rm surv}$ of the initially capable binaries that survive unperturbed long enough to undergo AGB-MT by age $t_{\rm evol}=11\,{\rm Gyr}$. Of these, we only count stars enriched by AGB-MT that are subgiants or RGB stars at $t_{\rm evol}$, like most detected AGB-MT stars. \emph{Right}: $f_{\rm surv}$ (black) decreases with increasing cluster half-mass density $\rho_{h}$, measured at $t_{\rm evol}$ for all GCs except the least dense. Since that model dissolves at $t\approx7\,{\rm Gyr}$, its plotted $\rho_h$ (measured at $t=6\,{\rm Gyr}$) is simply illustrative. The total binary fraction $f_{\rm b}$ for mass ratios $q>0.5$ (red) falls similarly with $\rho_h$, but less steeply as it includes binaries too tight to undergo AGB-MT without also undergoing common envelope. The red scatter points are the corresponding binary fractions $f_{b,{\rm obs}}(q>0.5)$ observed in a subset of Galactic GCs \cite[Table 1 of][excluding values flagged there as unreliable]{JiBregman2015}. 
The fraction of subgiant/RGB stars at $t_{\rm evol}$ that likely underwent AGB-MT (blue) is $0\%$ in all models, suggesting AGB-MT via standard RLOF alone is too rare to explain the observed AGB-MT stars in GCs.}
\label{fig:binary_survival}
\end{figure*}

\begin{figure*}[ht!]
\centering
\includegraphics[width=\textwidth]{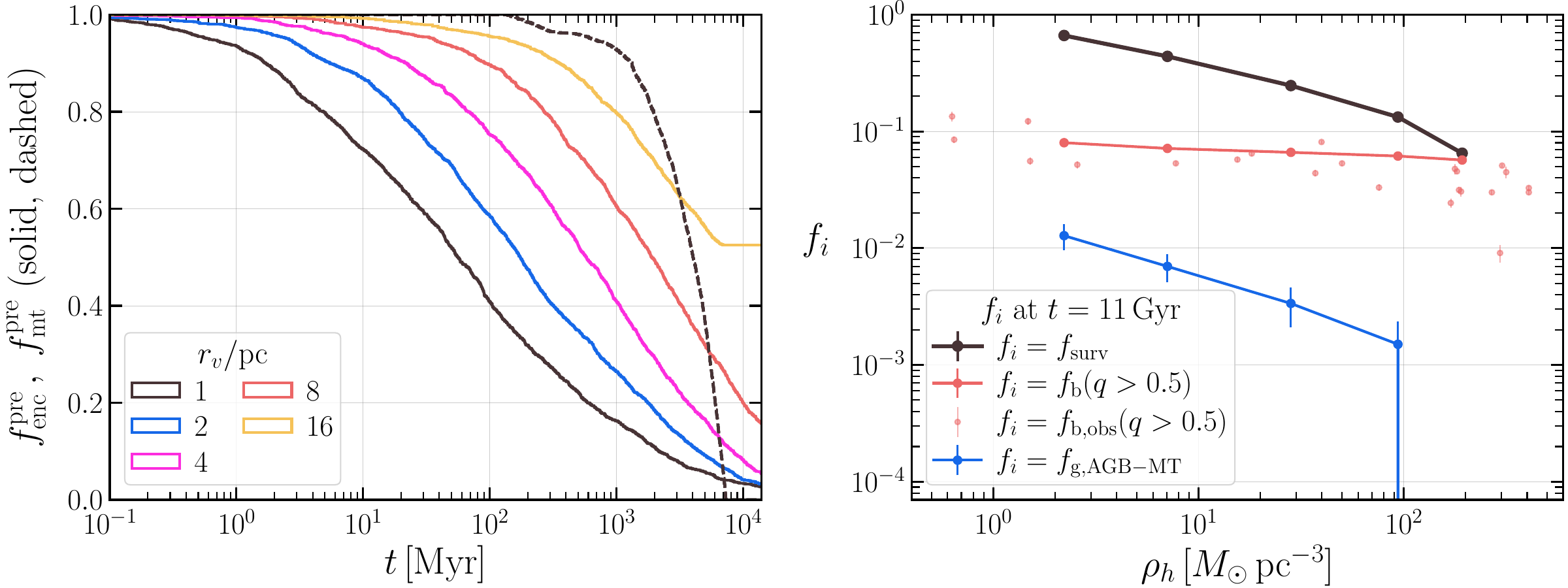}
\caption{Scenario 2: most AGB-MT occurs via WRLOF. Details mirror Figure~\ref{fig:binary_survival3}, with smoother curves in the left panel due to the larger sample of binaries. The right panel's vertical axis is also now logarithmic to show how the fraction of subgiant/RGB stars at $t_{\rm evol}$ that likely underwent AGB-MT is now nonzero and falls as $\rho_h$ rises. (Note the error bars are Poissonian and the curves would be nearly indistinguishable from Figure~\ref{fig:binary_survival} if shown in linear scale.) The mean $f_{\rm g,AGB-MT}$ in the intact GCs is $0.3\%$, consistent with the average from observed GCs \citep[$0.4\%$;][]{DOrazi2010}. The only model yielding $f_{\rm g,AGB-MT}>1\%$ is the fully-dissolved GC ($r_v=16\,{\rm pc}$), where the selection is from its stream.}
\label{fig:binary_survival3}
\end{figure*}

\section{Results} \label{sec:results}
Figure~\ref{fig:binary_survival} shows the results when assuming AGB-MT only arises from standard RLOF ($N_{\rm mt,iso}=135$). The left panel shows the $t_{\rm enc}$ and $t_{mt}$ distributions for each simulation---i.e., the time evolution in the fraction of binaries initially capable of AGB-MT that have yet to undergo a strongly perturbing encounter ($f^{\rm pre}_{\rm enc}$) or initiate mass-transfer ($f^{\rm pre}_{\rm mt}$; assuming left unperturbed). Since encounter rates grow with density, $f^{\rm pre}_{\rm enc}$ falls fastest at the lowest $r_v$. For each GC, the crossing point of $f^{\rm pre}_{\rm mt}$ and $f^{\rm pre}_{\rm enc}$ approximates the fraction $f_{\rm surv}$ (right panel's black curve) of the initially capable binaries that survive long enough to undergo AGB-MT. As expected, $f_{\rm surv}$ decreases steeply with increasing GC density---here, the mean density $\rho_h$ within the GC's half-mass radius at $t_{\rm evol}$ (or $6\,{\rm Gyr}$ for the least dense model since it dissolves before $t_{\rm evol}$).

The right panel similarly shows binary fraction $f_b$ for mass ratios $q>0.5$ in both the models (red curve) and observed GCs \citep[scattered red points; Table 1 of][]{JiBregman2015}. Both datasets exhibit the $f_b$--$\rho_h$ anticorrelation expected from dynamical binary hardening and disruption. Our initial conditions are not explicitly calibrated to match present-day $f_b$ in GCs, but it is reassuring that $f_b(q>0.5)\approx5\%$ both observationally and at $t_{\rm evol}$ in the models (despite starting at ${\approx}17\%$). This suggests the initial binary prescriptions are not too far off. At high $\rho_h$, the models somewhat overpredict observed $f_b(q>0.5)$. This discrepancy could reflect lower observational completeness at higher $\rho_h$ or arise from an initial period distribution that skews too low.

The right panel also indicates the fraction $f_{\rm g,AGB-MT}$ of subgiants or RGB stars at $t_{\rm evol}$ that are survivors from the binaries initially capable of AGB-MT.\footnote{This wording reflects limitations to \texttt{CMC} output---patched in an upcoming upgrade---that make a complete count of all AGB-MT stars impossible in post-processing. Our count excludes AGB-MT events in binaries formed dynamically after $t=0$, and assumes a ${<}10\%$ change in semi-major axis or eccentricity will never prevent AGB-MT in binaries otherwise on track for it. The former inaccuracy is larger so our estimate is conservative.}
Thus $f_{\rm g,AGB-MT}$ estimates the fraction of a GC's present (sub)giants that are AGB-MT stars. From ${\sim}10^4$ subgiants and RGB stars at $t_{\rm evol}$ across our simulations \emph{we find no likely AGB-MT stars resulting from standard RLOF}. This implies AGB-MT via standard RLOF alone is too rare to explain observed AGB-MT stars in GCs.

\begin{figure*}[ht!]
\centering
\includegraphics[width=\textwidth]{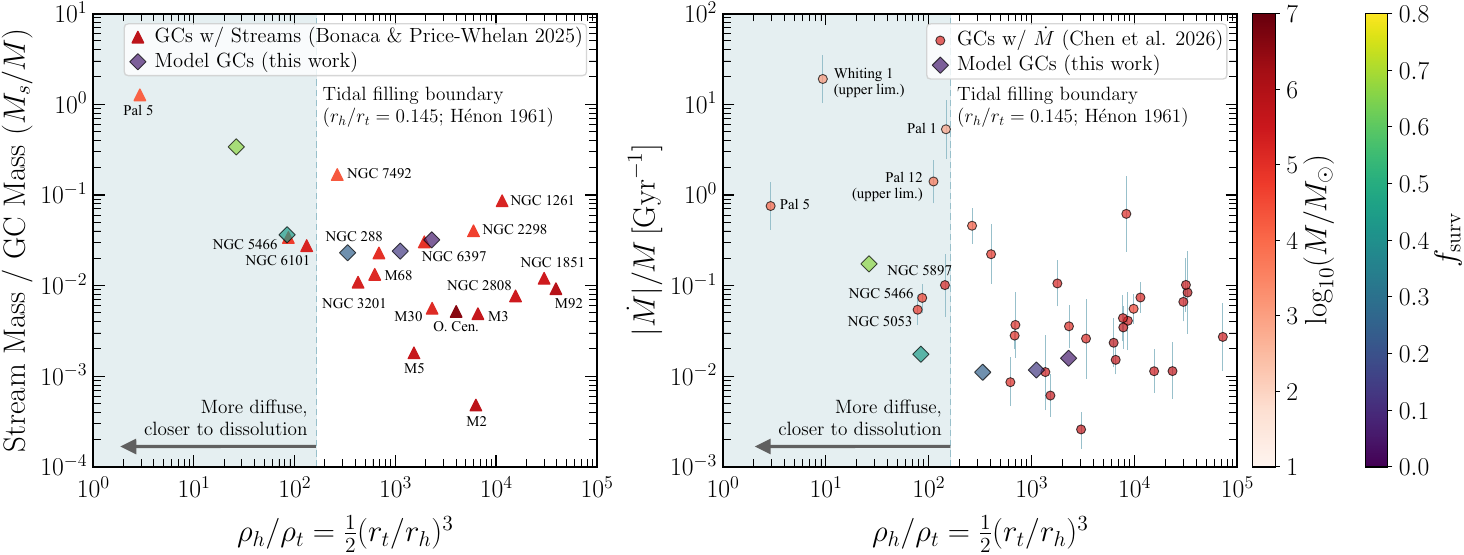}
\caption{Two measures of dissolution in our model GCs, and Galactic GCs with observed extratidal debris, versus a measure of tidal filling: the ratio between the GC's half-mass and tidal densities, $\rho_h/\rho_t=(1/2)(r_t/r_h)^3$.
\emph{Both panels:} Our models are represented by diamonds colored by AGB-MT binary survival fraction, $f_{\rm surv}$, while Galactic GCs (triangles or circles) are colored by their current mass, $M$. We use $M$, $r_h$, and $r_t$ measured by \citet[][2023 edition]{BaumgardtCatalog2020}. Note their $r_t$ accounts for the eccentricity of each GC's orbit \citep[via Equation~8 of][]{Webb2013}. The gray area encompasses GCs that are tidally overfilling \citep[$r_h/r_t\geq0.145$;][]{Henon1961}, which are most vulnerable to tides and closest to dissolution. The model GCs born least dense and with highest $f_{\rm surv}$ expanded to be tidally overfilling or close to it, so evaporate faster with \emph{decreasing} density. Those that remain tidally underfilling instead evaporate slightly faster with \emph{increasing} density, due to faster relaxation.
\emph{Left panel}: Lower bound on the ratio between the stream mass and GC mass ($M_s/M$) in our models and 18 GCs with observed streams. For the observed GCs, we use $M_s$ from two surveys \citep{Shipp2018, Ibata2024} collated in Table~C.2 of \cite{BonacaPriceWhelan2025}. Both surveys extrapolated $M_s$ from the number of identified stream members assuming a stellar initial mass function and a magnitude-complete sample; as sample purity was prioritized over completeness, the reported $M_s$ is a lower bound. For our GC models, we use $M_s$ and $M$ at $11\,{\rm Gyr}$ (or $6\,{\rm Gyr}$ for the dissolved GC). To mimic observing limits, we determine model $M_s$ counting only stars that escaped the GC in the preceding $2\,{\rm Gyr}$---the typical temporal extent of observed streams in terms of time since escape \citep[e.g.,][]{Chandra2026}.
\emph{Right panel:} The specific mass-loss rate, $\lvert\dot{M}\rvert/M$ in our models and 30 GCs with $\dot{M}$ measurements based on their extratidal debris, each GC's stellar mass function, and the observational selection function \citep[][their Table~1]{Chen2026}. For our GC models, we use the mean $\lvert\dot{M}/M\rvert$ over the preceding $1\,{\rm Gyr}$, measured at the same GC age as in the left panel. 
}
\label{fig:Mdot}
\end{figure*}

Figure~\ref{fig:binary_survival3} shows the results when assuming WRLOF contributes to AGB-MT. Since WRLOF allows AGB-MT to occur when the donor's radius is several times smaller than its Roche lobe, accounting for WRLOF leads to ${>}10$ times more binaries initially capable of AGB-MT if left unperturbed ($N_{\rm mt,iso}=1813$ versus $135$, respectively). This is enough to make $f_{\rm g,AGB-MT}$ non-zero (right panel, blue cruve), and the $0.3\%$ mean across all models is consistent with the mean from observed GCs ($0.4\%$). So AGB-MT via WRLOF is sufficient to explain observed rates of AGB-MT stars in GCs.

As expected, $f_{\rm g,AGB-MT}$ also falls monotonically with rising $\rho_h$, making the fully dissolved GC the only model with $f_{\rm g,AGB-MT}>1\%$. The selection in this case is from the GC's stellar ejecta in the last $2\,{\rm Gyr}$ of its life, representative of the densest part of its stream. Yet even its $f_{\rm g,AGB-MT}\approx1.3\%$ may be several times lower than the mean in observed streams (${\approx}4\%$ but with low sample size). So while cluster dynamics may help explain why streams have higher $f_{\rm g,AGB-MT}$ than intact GCs, it is possible other phenomena or selection effects may share responsibility for any observed discrepancy.

To help extrapolate to stellar streams, Figure~\ref{fig:Mdot} shows the ratio between stream mass and GC mass ($M_s/M$; left panel) and the GC mass-loss rate ($\lvert\dot{M}\rvert/M$; right panel) versus a measure of how tidally filling a GC is: the ratio between its half-mass density and tidal density, $\rho_h/\rho_t=(1/2)(r_t/r_h)^3$, where $r_h$ and $r_t$ are the half-mass and tidal radii, respectively. Our GC models appear in each panel as diamonds colored by $f_{\rm surv}$ while observed GCs with known streams \citep[Table~C.2 of][]{BonacaPriceWhelan2025} or evidence of extratidal debris \cite[Table~1 of][]{Chen2026} appear as triangles and circles colored by GC mass in the left and right panels, respectively. While subject to large observational uncertainties (e.g., completeness), these metrics of GC dissolution are broadly consistent with each other. Tidally overfilling GCs \citep[furthest left, formally occupying the gray area with $r_h/r_t>0.145$;][]{Henon1961} are the closest to dissolution. Streams from these GCs, including the classic examples of Palomar~5 \citep{Odenkirchen2001} and NGC~5466 \citep{Belokurov2006, Grillmair2006}, are preferentially dense. This is a result of the $\dot{M}/M$ rising with falling $\rho_h/\rho_t$ within and just beyond the tidally overfilling regime---i.e., until a shallow turnaround in the trend at higher $\rho_h/\rho_t$, where relaxation starts to drive dissolution over tides (Section~\ref{sec:probing_progenitor_density}). That $f_{\rm surv}$ also increases with $\lvert\dot{M}/M\rvert$ in this regime further suggests AGB-MT remnants should be more common in stellar streams than dense GCs. 

\section{Discussion} \label{sec:discussion}

\subsection{Modeling Considerations}
While the trends in our results are clear and well-founded physically, the absolute $f_{\rm AGB,MT}$ predictions should be taken as approximate. Absolute fractions of binaries or binary evolution products like AGB-MT stars depend on the primordial binary population and mass transfer physics, which are both quite uncertain. WRLOF, in particular, is not considered in the stellar evolution prescriptions built into most star cluster modeling codes, including our own. We thus used a simple post-processing criterion to determine \textit{onset} of WRLOF only, without explicitly conducting any mass transfer that could otherwise alter the product's detectability (e.g., by making it too massive to be a giant at present) or shrink the binary separation enough for common envelope to occur, preventing detectable chemical enrichment. Conservatively, we also did not consider the contribution to AGB-MT star production from binaries formed dynamically (e.g., exchange interactions or three-body binary formation) or the possibility that some strongly perturbed binaries may still manage to undergo AGB-MT. Additional AGB-MT from these channels is likely small but not negligible, meriting further investigation in future work. For all these reasons, the \textit{relative} abundance of AGB-MT stars between streams or GCs should be more robust indicators of GC density than absolute abundances.

\subsection{Probing Progenitor Density} \label{sec:probing_progenitor_density}
Our results suggest that the abundance of AGB-MT stars in a stellar stream can help constrain the density of its progenitor, which is hard to infer from stream morphology or kinematics alone. Stream kinematics are sensitive to perturbative heating from Galactic substructure, and while stream density tracks the progenitor's mass-loss rate $\dot{M}$, this scales non-monotonically with $\rho_h$ for a fixed GC mass and orbit (e.g., Figure~\ref{fig:Mdot}). For compact, tidally underfilling GCs, which are still expanding, $\dot{M}$ rises slightly with density due to faster relaxation \citep{BinneyTremaine2008, McLaughlin2008, Weatherford2023}. This regime is typical of most Galactic GCs at present \citep{Gieles2008}. Once a GC expands enough to overflow its tidal boundary, however, $\dot{M}$'s dependence on density becomes either negligible \citep{Henon1961, Gieles2011}, or entirely reverses if the GC retains a black hole population that can drive further expansion \citep{Merritt2004, Mackey2008, BreenHeggie2013, CMCCatalog}. Indeed, there is evidence of dynamically significant black hole retention in many Galactic GCs \citep{Weatherford2020}. In this tidally overfilling regime, expediently accessed by GCs born especially diffuse, $\dot{M}$ instead rises with decreasing density \citep{Giersz2019, Gieles2023, Roberts2025}. A similarly prominent stream can result from either regime (either side of the minimum $\dot{M}/M$ in Figure~\ref{fig:Mdot}).

Unlike $\dot{M}$, the abundance of binaries and their stellar evolution products varies monotonically with GC density, as their survival or formation depends strongly on encounter rates. With $\dot{M}$ inferred from stream density, the abundance of such products, including AGB-MT stars, could be used to break the above degeneracy by placing progenitors on the dense or diffuse branch of Figure~\ref{fig:Mdot}. Binaries overall are especially useful, being more common than any specific product of binary evolution; $f_b(q>0.5)$ falls with $\rho_h$ in both our model GCs/streams and observed GCs (Figure~\ref{fig:binary_survival}), similar to recent results from direct $N$-body models \citep{Phillips2026}. Collision products in streams (e.g., blue stragglers) may also trace progenitor density.

However, binaries are heavier than typical stars, delaying their escape as mass segregation makes them less vulnerable to tidal stripping and their greater inertia reduces the kicks they receive in encounters \citep{Weatherford2023}. Since tidally overfilling or black hole-rich GCs lose mass fastest at the ends of their lives, the delay can concentrate binaries in a stream's densest, most detectable region \citep{Balbinot2018, Gieles2023, Phillips2026}. Future work to constrain progenitor densities from binaries and mass-transfer products in streams must account for these complexities.

\subsection{Observational Considerations}
AGB-MT stars are rare; distinguishing $f_{\rm g,AGB-MT}\approx 1\%$ in a dense GC requires substantially larger samples than the ${<}20$ stars studied spectroscopically per stream so far. Fractions pooled across streams from existing surveys such as S$^5$ \citep{Li2019_S5} could provide initial constraints on the typical densities of dissolved progenitors, but robust stream-by-stream inference will require larger samples extending below the main-sequence turnoff. High-resolution spectroscopy will also remain essential to identify AGB-MT stars in streams via their detailed chemical abundances \citep[e.g.,][]{Ji2020b, Usman2024}.

Another complexity is the metallicity dependence of AGB-MT signatures, reflecting both AGB nucleosynthetic yields and mass-transfer efficiency. Though the overall CEMP fraction rises with falling metallicity, reaching ${\approx}43\%$ at ${[\rm Fe/H]}\leq-3.0$, the CEMP-$s$ fraction associated with AGB-MT instead declines from ${\approx}13\%$ at ${[\rm Fe/H]}<-2.0$ to ${\approx}5\%$ at ${[\rm Fe/H]}<-3.0$ \citep{Placco+2014, Arentsen+2022}. Large samples from Gaia XP similarly reveal strong metallicity and environmental dependencies, including a lower incidence of carbon-rich stars in GCs than in the field, consistent with differences in their binary populations \cite[e.g.,][finding only two carbon-rich stars out of a sample of ${\approx}1800$ GC stars]{ArdernArentsen+2025}. Probing these dependencies will be essential to disentangle the effects of binary evolution from cluster dynamics.

\subsection{Emerging Observational Opportunities}
The next decade of astronomy is set to usher in a ``binary star renaissance’’ \citep{ElBadry2024a}, transforming our understanding of binary demographics alongside the dynamical evolution and dissolution of GCs. Gaia Data Release 4 will provide epoch astrometry and improved non-single-star solutions, enabling the identification and characterization of a much wider variety of binaries \cite[e.g.,][]{ElBadry2024b,Guerriero+2026}. The Nancy Grace Roman Space Telescope will further extend time-domain studies into fainter and more crowded stellar environments, complementing upcoming Gaia data releases and providing new constraints on binary populations, stellar proper motions, and the morphology of stellar streams \citep{Gandhi+2023,Grunblatt+2023,Aganze+2024}.

Upcoming ground-based spectroscopic surveys will provide a complementary avenue for identifying AGB-MT stars. The Via spectroscopic survey of stellar streams \citep{ViaCollab} will obtain spectra at resolution $R=15,000$ over $505$–$595\,{\rm nm}$, encompassing Mg and Na tracers of internal GC populations alongside $\mathrm{C}_2$ Swan bands and Ba{\sc ii} as key tracers of AGB-MT. Similarly, 4MOST will provide spectra of comparable resolution across a broader wavelength range, enabling a richer chemical characterization including multiple $s$-process elements \citep{deJong+2019}. Together, these surveys will help distinguish intrinsic GC enrichment from AGB-MT signatures, while repeated radial velocity measurements will independently probe binarity.

Together, upcoming space-based astrometry and ground-based spectroscopy will provide an increasingly complete view of binary populations in both intact and dissolved GCs. These observations will further clarify how the abundance of AGB-MT systems varies with GC density, providing a new avenue to probe the dynamical history of dissolved GCs, including rates of stellar exotica and transient phenomena.

\section{Conclusion} \label{sec:conclusion}
In this paper, we have explored a potential dynamical explanation for why AGB-MT stars may be more abundant in GC stellar streams than intact GCs. To do so, we have examined the survival of relevant mass-transferring binaries in simulated GCs using initial binary prescriptions consistent with observations of binaries in the solar neighborhood and nearby star-forming regions. Our simulations also start with typical initial GC mass (kept constant for simplicity) but varying densities; over a typical GC lifetime, these clusters expand to cover well the range of observed half-mass densities in intact Milky Way GCs. We find the following:
\begin{itemize}
    \setlength\itemsep{-0.2em}
    \item At an $11\,{\rm Gyr}$ age typical of Galactic GCs, our models reproduce well their observed binary fractions (${\approx}5\%$, despite starting at ${\approx}17\%$). While there is some deviation at high GC density, this is an encouraging test of \texttt{CMC}'s latest initial binary prescriptions.
    \item The number of initial binaries capable of AGB-MT that survive GC dynamics to undergo AGB-MT strongly declines with increasing half-mass density.
    \item Standard RLOF without common envelope is too rare to account for the observed fraction of subgiant and RGB stars in GCs with signs of AGB-MT ($f_{\rm AGB,MT}\approx0.4\%$). Yet our models closely reproduce that fraction when accounting for potential WRLOF events, increasing AGB-MT star yields by a factor ${>}10$ and leading to a mean $f_{\rm AGB,MT}\approx0.3\%$ at $11\,{\rm Gyr}$.
    \item $f_{\rm AGB,MT}$ rises monotonically with falling density, from $0\%$ in our densest GC to $1.3\%$ in the stream of our least-dense GC, which fully dissolved. While that is below the ${\approx}4\%$ observed in stellar streams so far (from low sample sizes), GCs born even less dense or less massive could yield higher $f_{\rm AGB,MT}$. The trend at least indicates stellar streams should be more abundant in AGB-MT stars than most intact GCs.
\end{itemize}

The next decade of observational campaigns will provide the statistical information needed to establish robust rate estimates of binaries and AGB-MT stars in GCs and streams, accounting for magnitude limits and survey completeness, and to determine how their incidence varies with metallicity and birth environment. In parallel, further improvements in modeling mass transfer and stellar stream formation will enable increasingly direct comparisons between these observations and theoretical predictions. Together, such advances will contribute new paths to probe the evolution, dynamics, and structural properties of dissolved GCs from the stellar debris they leave behind. Such efforts will help explain chemically peculiar populations in the Galactic halo, inform constraints on dynamical production of stellar exotica and transients, and improve stellar stream modeling necessary for accurate stream-based dark matter inference.

\vspace{-0.5cm}
\begin{acknowledgements}
We thank Ana Bonaca, Sergey Koposov, Kyle Kremer, Fred Rasio, and Andrew Robertson for valuable discussions that improved this work. 
S.A.U. thanks Alex Ji for support during this project.
Support for this work was provided by NASA through the NASA Hubble Fellowship grant \#HST-HF2-51560 awarded by the Space Telescope Science Institute, which is operated by the Association of Universities for Research in Astronomy, Inc., for NASA, under contract NAS5-26555.
\end{acknowledgements}

\begin{contribution}
N.C.W. led the paper write-up and contributed the modeling and data analysis. M.M. and S.A.U. helped guide criteria for determining AGB-MT stars and contributed to the discussion of AGB-star mass transfer physics. S.A.U.'s observational work (covered in other papers in prep.) inspired the project.
\vspace{0.3cm}
\end{contribution}

\software{
\code{CMC} \citep{CMC_v1.0},
\code{COSMIC} \citep{COSMIC},
\code{Astropy} \citep{Astropy, Astropy2, Astropy3},
\code{matplotlib} \citep{Matplotlib},
\code{SciPy} \citep{SciPy},
\code{NumPy} \citep{NumPy},
\texttt{pandas} \citep{Pandas}.
}

\clearpage
\bibliographystyle{aasjournalv7.1}
\bibliography{main}
\end{document}